\documentclass[preprint,aps,showpacs,preprintnumbers,amsmath,amssymb]{revtex4}

\usepackage{graphicx}
\usepackage{dcolumn}
\usepackage{bm}

\begin{document}


\title{Multiphoton Femtosecond Control of Resonance-Mediated Generation of Short-Wavelength Coherent Broadband Radiation}
\author{Leonid Rybak, Lev Chuntonov, Andrey Gandman, Naser Shakour}
\author{Zohar Amitay}
\email{amitayz@tx.technion.ac.il}
\affiliation{
Schulich Faculty of Chemistry, Technion - Israel Institute of Technology, Haifa 32000, Israel}
\begin{abstract}
We introduce a new scheme for generating short-wavelength coherent broadband radiation
with well-controlled spectral characteristics.
It is based on shaping long-wavelength femtosecond pulse to coherently control
atomic resonance-mediated (2+1) three-photon excitation to a broad far-from-resonance continuum.
Here, the spectrum (central frequency and bandwidth)
of deep-ultraviolet coherent broadband radiation generated in Na vapor is experimentally controlled by
tuning the linear chirp we apply to the driving phase-shaped near-infrared femtosecond pulse.
This is a first step in implementing the full scheme for producing shaped femtosecond pulses
at wavelengths down-to the vacuum-ultraviolet range.
\end{abstract}
%
%

\maketitle

Femtosecond coherent control is a field 
of both scientific and technological importance
concerning a variety of physical and chemical photo-induced processes \cite{dantus_exp_review1,gerber_condensed_phase_control_review}.
It utilizes the broad coherent spectrum of femtosecond pulses, which provides a manifold of excitation pathways,
to control the photo-induced 
dynamics and outcome in 
ways that other means cannot.
The control handle is the pulse shape that is 
determined by the spectral characteristics of the broadband pulse, i.e., 
the amplitude, phase, and polarization of its different frequency components \cite{pulse_shaping}.  

Indeed,
over the past decade, coherent 
control using shaped femtosecond pulses
has been successfully demonstrated in the laboratory 
with many systems and processes \cite{dantus_exp_review1,gerber_condensed_phase_control_review}.
However, although very successful, these 
experiments have been limited to
atomic and molecular species that absorb 
visible (VIS) or near-infrared (NIR) light.
The reason is the experimental difficulties in shaping femtosecond pulses outside these spectral regions, where
the devices commonly-used for flexible pulse shaping
(liquid-crystal modulators 
and TeO$_{2}$ acousto-optic modulators) \cite{pulse_shaping}    
are inappropriate. 
Extending the techniques to generate shaped 
pulses at short wavelengths down to the deep-ultraviolet (UV) 
and vacuum-ultraviolet (VUV) regions
will significantly extend the variety of 
molecules 
accessible for coherent control,
since most electronic transitions are in this spectral range. 


Recently, significant progress has been made in producing shaped UV pulses 
based on generating the UV pulse from a source NIR femtosecond pulse
via (nonresonant) nonlinear optical processes in crystals
\cite{Sauerbrey_UV_400nmSHG_shaping_idirect,Sauerbrey_UV_200nm_shaping_idirect,
Weiner_UV_400nm_and_ideas_shaping_indirect,Midorikawa1_UV_shaping_indirect,Gerber_UV_400nmSHG_shaping_indirect_exps,Riedle_UV_shaping_idirect,
Midorikawa2_UV_shaping_indirect,Brixner_Gerber_UV_shaping_indirect,Brixner_Gerber_UV_400nm_polarization_shaping_indirect,
Motzkus_Sauerbrey_UV_shaping_directMicromirror,Rabitz_UV_400nm_shaping_directAOM,
Dogariu_UV_400nm_shaping_directAOM,Weinacht_UV_shaping_directAOM}.
%
%
For the UV pulse shaping the corresponding works have followed two approaches. 
One approach is of direct UV shaping
\cite{Motzkus_Sauerbrey_UV_shaping_directMicromirror,Rabitz_UV_400nm_shaping_directAOM,
Dogariu_UV_400nm_shaping_directAOM,Weinacht_UV_shaping_directAOM},
where the generated UV pulse undergoes direct shaping using
new types of special devices that are appropriate to the UV spectral range.
%
%
%
The other approach is of indirect UV shaping
\cite{Sauerbrey_UV_400nmSHG_shaping_idirect,Sauerbrey_UV_200nm_shaping_idirect,
Weiner_UV_400nm_and_ideas_shaping_indirect,Midorikawa1_UV_shaping_indirect,Gerber_UV_400nmSHG_shaping_indirect_exps,Riedle_UV_shaping_idirect,
Midorikawa2_UV_shaping_indirect,Brixner_Gerber_UV_shaping_indirect,Brixner_Gerber_UV_400nm_polarization_shaping_indirect},
where the NIR-to-UV nonlinear optical conversion is done with a (conventionally)
shaped NIR pulse and, hence, the generated UV pulse is already a shaped one.
%
%
The output shaped broadband UV spectrum is determined by the input shaped NIR spectrum following its 
NIR-to-UV nonlinear optical transformation. 
%
%
%
However, even though all these studies 
provide a feasible pathway to UV pulse shaping,
their basic use of optical crystals 
actually limits the generated pulse to UV    
wavelengths above 200-250~nm. 
%
%


In this Letter, we pursue a line of action that is generally applicable
for generating short-wavelength shaped pulses down-to the VUV spectral range. 
%
%
It follows the above-mentioned indirect shaping approach, 
however, instead of using optical crystals, the shaped short-wavelength pulse is generated
from a shaped long-wavelength pulse
via an atomic resonance-mediated (2+1) three-photon excitation to a broad far-from-resonance continuum.
%
The resulting transient 
polarization drives 
the generation of the shaped short-wavelength coherent broadband radiation.
%
Resonance-mediated three-photon excitation in atomic vapors has actually been employed 
very successfully in the past for efficient and tunable generation  
of short-wavelength radiation, however it has been used only for
generating narrowband (nanosecond or picosecond) UV and VUV coherent radiation by narrowband NIR and VIS excitation
(for example, see \cite{Harris_THG,narrow_VUV1,narrow_VUV2,narrow_VUV3,narrow_VUV4,narrow_VUV5,narrow_VUV6,
narrow_VUV7,narrow_VUV8,narrow_VUV9,narrow_VUV10_eden_review2});
The present broadband nature introduces a completely new dimension to the process.
%
%
Our new scheme also utilizes previous works on femtosecond broadband coherent control of resonance-mediated (2+1) three-photon
absorption \cite{Amitay_3ph_2plus1_2,Amitay_selective_2ph_3ph},
extending them from a case of a single (real) final state to a case of
a continuous manifold of multiple (virtual) final states.
Here, as a first step in implementing the full scheme,
the spectrum (central frequency and bandwidth) of deep-ultraviolet coherent broadband radiation
generated in Na vapor is experimentally controlled by
tuning the linear chirp that we apply to the driving phase-shaped near-infrared femtosecond pulse.

\begin{figure} [htbp]
\includegraphics[scale=0.65]{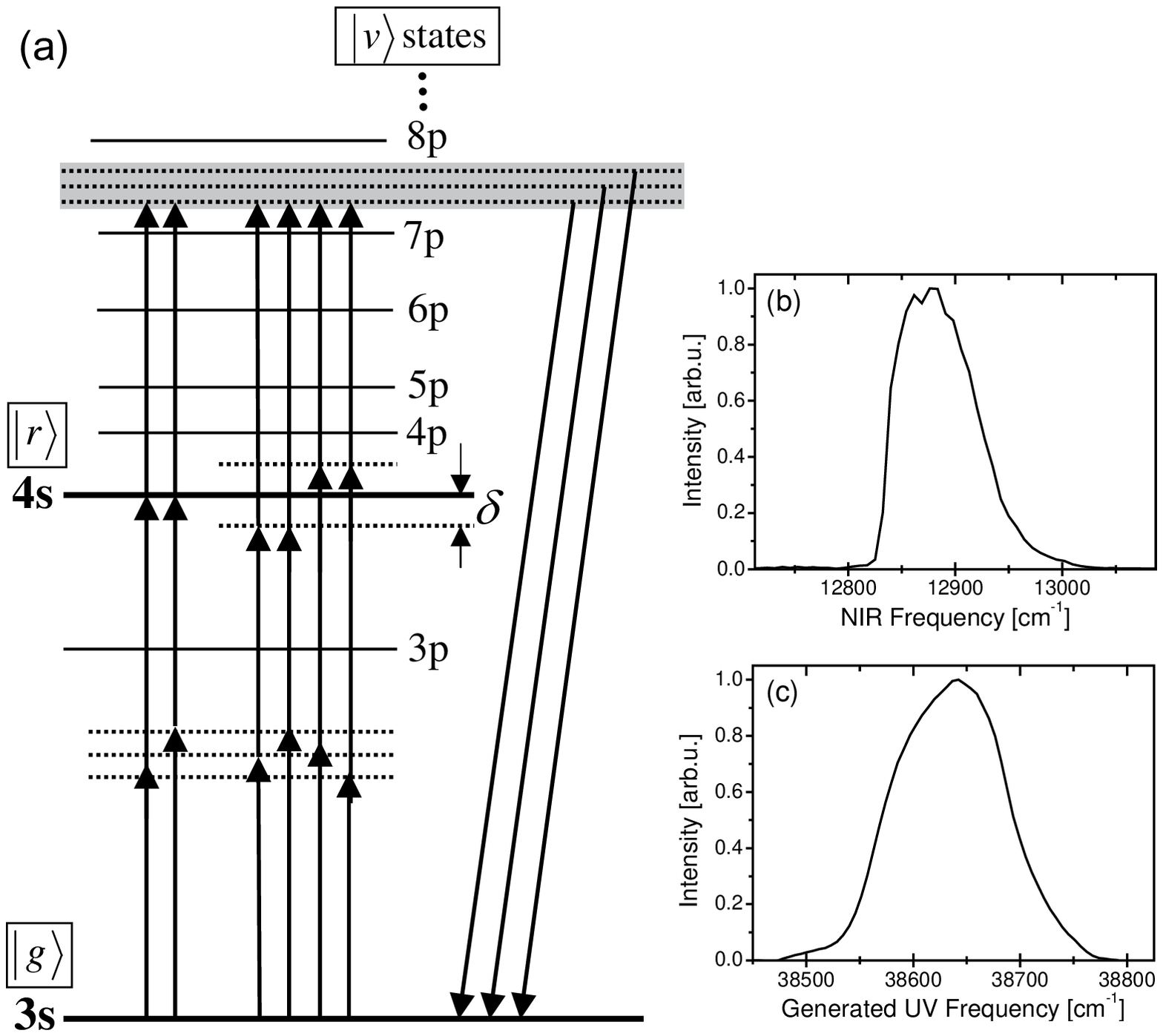}
\caption{(a) The excitation scheme of Na for the generation of
deep-ultraviolet (UV) coherent broadband radiation by shaped
near-infrared (NIR) femtosecond pulses via resonance-mediated (2+1)
three-photon excitation to a broad far-from-resonance continuum. (b)
The spectrum of the driving NIR femtosecond pulses used here. (c)
The measured spectrum of the UV coherent broadband radiation
generated in Na by the NIR transform-limited pulse.} \label{fig_1}
\end{figure}

Figure~\ref{fig_1}(a) shows the atomic broadband 
excitation scheme considered here, 
%
involving an initial ground state $\left|g\right\rangle$ and an excited state $\left|r\right\rangle$
of one symmetry and a manifold of excited states $\left|v\right\rangle$ of another symmetry.
Irradiation with a 
shaped long-wavelength femtosecond pulse 
induces a resonance-mediated (2+1) three-photon excitation from $\left|g\right\rangle$
via $\left|r\right\rangle$ to
a broad far-from-resonance continuum, 
resulting in a time-dependent atomic polarization that drives
the generation of a corresponding shaped short-wavelength coherent broadband radiation.
The last coherent emission step is associated with a corresponding de-excitation to $\left|g\right\rangle$.
Along this whole photo-induced 
process the $\left|v\right\rangle$ states are accessed nonresonantly   
and provide the $\left|g\right\rangle$-$\left|r\right\rangle$ nonresonant two-photon couplings
at the one- and three-photon excitation energies.

Expressing the temporal 
atomic polarization 
using third-order time-dependent perturbation theory
and transforming it to the frequency domain,   
we have obtained the generated
short-wavelength spectral 
field $E_{out}(\Omega)$ 
at frequency $\Omega$ to be given by
\begin{eqnarray}
E_{out}(\Omega) & \propto & \mu^{2}_{g,r} \chi^{2}_{r,g}
\left[A^{(2+1)on-res}(\Omega) + A^{(2+1)near-res}(\Omega)\right] \; ,
\label{eq1} \\
A^{(2+1)on-res}(\Omega) & = & i \pi A^{(2)}(\omega_{r,g}) E(\Omega-\omega_{r,g}) \; ,
\label{eq2} \\
A^{(2+1)near-res}(\Omega) & = & -\wp\int_{-\infty}^{\infty}\frac{1}{\delta}A^{(2)}(\omega_{r,g}-\delta)E(\Omega-\omega_{r,g}+\delta)d\delta \; ,
\label{eq3} \\
A^{(2)}(\xi) & = & \int_{-\infty}^{\infty} E(\omega)E(\xi-\omega)d\omega \; , 
\label{eq4}
\end{eqnarray}
%
%
with $\omega_{r,g}$ being the $\left|g\right\rangle$-$\left|r\right\rangle$ transition frequency.
The $E(\omega)$$\equiv$$\left|E(\omega)\right|\exp{[i\Phi(\omega)]}$ is the long-wavelength spectral field of the excitation pulse,
with $\left|E(\omega)\right|$ and $\Phi(\omega)$ being, respectively,
the spectral amplitude and phase of frequency $\omega$.
For the (unshaped) transform-limited (TL) pulse, 
$\Phi(\omega)$=0. 
%
The $\mu^{2}_{r,g}$ and~$\chi^{2}_{g,r}$ are
the $\left|g\right\rangle$-$\left|r\right\rangle$ nonresonant
two-photon couplings (due to the $\left|v\right\rangle$ states)
at 
the $\omega_{c}$ and $\omega_{r,g}+\omega_{c}$ 
excitation regions, with $\omega_{c}$ being the carrier frequency of $E(\omega)$. 
%
%
%
The spectral intensity generated at $\Omega$ is 
$I_{out}(\Omega)\propto\left|E_{out}(\Omega)\right|^{2}$, and
the total 
short-wavelength yield is 
$Y_{out}=\int_{-\infty}^{\infty}I_{out}(\Omega)d\Omega$.

As can be concluded from the above frequency-domain
formulation and is illustrated in Fig.~\ref{fig_1}(a),
the short-wavelength 
field $E_{out}(\Omega)$ generated at a frequency $\Omega$
results from the interferences among all the possible resonance-mediated three-photon excitation pathways
from $\left|g\right\rangle$ to 
the excitation energy corresponding to $\Omega$.
%
Each of these pathways is either on 
or near resonance with the intermediate state $\left|r\right\rangle$.
A pathway of detuning $\delta$
involves a nonresonant absorption of two photons with a frequency sum of $\omega_{r,g}$$-$$\delta$
and the absorption of a third complementary photon of frequency $\Omega$$-$$(\omega_{r,g}$$-$$\delta)$.
%
%
The on- ($\delta$=0) and near-resonant ($\delta$$\ne$0) pathways reaching the energy of $\Omega$ are interfered separately in
$A^{(2+1)on-res}(\Omega)$ and $A^{(2+1)near-res}(\Omega)$
($\wp$ is the Cauchy principal value).
They 
are expressed using the parameterized amplitude $A^{(2)}(\xi)$
interfering all the nonresonant two-photon pathways of frequency sum $\xi$, 
each composed of photon pair $\omega$ and $\xi$$-$$\omega$. 
As the exciting long-wavelength field $E(\omega)$ 
controls the amplitudes and phases of the different interfering three-photon pathways,  
it also controls the amplitude and phase of
the short-wavelength field $E_{out}(\Omega)$ generated at each~$\Omega$.

The model system of the study is the sodium (Na) atom,
with the $3s$ ground state as $\left|g\right\rangle$, the $4s$ state as $\left|r\right\rangle$,
and the manifold of $p$ states as $\left|v\right\rangle$ [see Fig.~\ref{fig_1}(a)].
The $3s$-$4s$ transition frequency is $\omega_{r,g} \equiv \omega_{4s,3s} = 25740$~cm$^{-1}$ corresponding to two 777-nm photons.
The sodium is irradiated with phase-shaped linearly-polarized near-infrared (NIR) femtosecond pulses having
a spectrum peaked at $\omega_{0}$=12876~cm$^{-1}$ (776.6~nm) with 88-cm$^{-1}$ bandwidth (FWHM), 
as shown in Fig.~\ref{fig_1}(b).
The pulse energy is 10~$\mu$J
corresponding to a TL peak intensity of $I_{TL}$$\sim$3$\times$10$^{10}$~W/cm$^{2}$.
The NIR spectrum results from blocking an original spectrum at its lower-frequency edge
to prevent three-photon resonant access to the $7p$ state
[$\omega_{7p,4s} = 12801$~cm$^{-1}$ (781.2~nm)],
which is undesired for generating short-wavelength
radiation of broadband nature.
The interaction with the NIR excitation pulse leads to the generation of
a coherent broadband UV radiation. 
Experimentally, a sodium vapor in a heated cell at 900 K 
with Ar buffer gas is irradiated with the NIR laser pulses after
they undergo shaping in a setup incorporating a pixelated liquid-crystal spatial light phase modulator \cite{pulse_shaping}.
The shaping resolution is 2.05~cm$^{-1}$. 
The coherent UV radiation emitted in the propagation direction of the NIR beam is measured
using a UV-spectrometer coupled to a time-gated camera system.
The corresponding overall UV spectral resolution is 35~cm$^{-1}$~(0.23~nm),     
with 5.8-cm$^{-1}$ spectral width of each camera pixel.

Figure~\ref{fig_1}(c) shows the measured coherent UV spectrum generated by the NIR TL pulse.
As seen, it is of broadband nature with a 
bandwidth of 125~cm$^{-1}$ around its maximum at $\Omega$=38642~cm$^{-1}$ (258.78~nm).
After deconvoluting the measurement 
resolution, the UV spectral bandwidth amounts to 
120~cm$^{-1}$ ($\sim$125-fs TL duration). 
The peak UV frequency of 38642~cm$^{-1}$
corresponds to $\omega_{4s,3s}$ plus a frequency of 12902~cm$^{-1}$  
that is around the peak NIR frequency $\omega_{0}$.
As seen from Eqs.~(\ref{eq2})-(\ref{eq4}),
the on-resonant amplitude $A^{(2+1)on-res}(\Omega)$ is by-definition maximal at $\Omega$=$\omega_{4s,3s}$+$\omega_{0}$,
while the peak location of the near-resonant amplitude $A^{(2+1)near-res}(\Omega)$
can deviates from it according to the specific NIR excitation spectrum.

In terms of the absolute efficiency of the UV generation process,
under our experimental conditions,~at a temperature of 900 K corresponding to 
Na density of 3.7$\times$10$^{17}$~cm$^{3}$,
the total energy of the emitted coherent broadband UV radiation reaches 
about~1\% of the 
NIR pulse energy for the TL excitation of $I_{TL}$$\sim$3$\times$10$^{10}$~W/cm$^{2}$. 
As expected for a coherent generation process,
the UV generation efficiency exhibits a quadratic dependence ($n_\textrm{Na}^{2}$)
on the Na density $n_{\textrm{Na}}$
(controlled by the cell temperature).
Up to the NIR TL peak intensity of $I_{TL}$$\sim$3$\times$10$^{10}$~W/cm$^{2}$
the UV generation efficiency exhibits a cubic dependence ($I_{TL}^{3}$) on $I_{TL}$,
indicating the third-order perturbative regime associated with Eqs.~(\ref{eq1})-(\ref{eq4}).
Higher values of $I_{TL}$ 
lead to a deviation from this regime,
which we avoid
in order to maintain~the
validity of the relatively simple theoretical description.

As a first control study on resonance-mediated generation of short-wavelength coherent broadband radiation,
we demonstrate control over the UV generation with
linearly-chirped NIR pulses shaped with
quadratic spectral phase patterns $\Phi(\omega)=\frac{1}{2} \kappa_{chirp}(\omega-\omega_{0})^{2}$
of variable chirp parameter $\kappa_{chirp}$.
The $\kappa_{chirp}$ value is 
tuned here from negative to positive values, with the zero value $\kappa_{chirp}$=0 corresponding to
the NIR TL pulse.
%
%

\begin{figure} [htbp]
\includegraphics[scale=0.65]{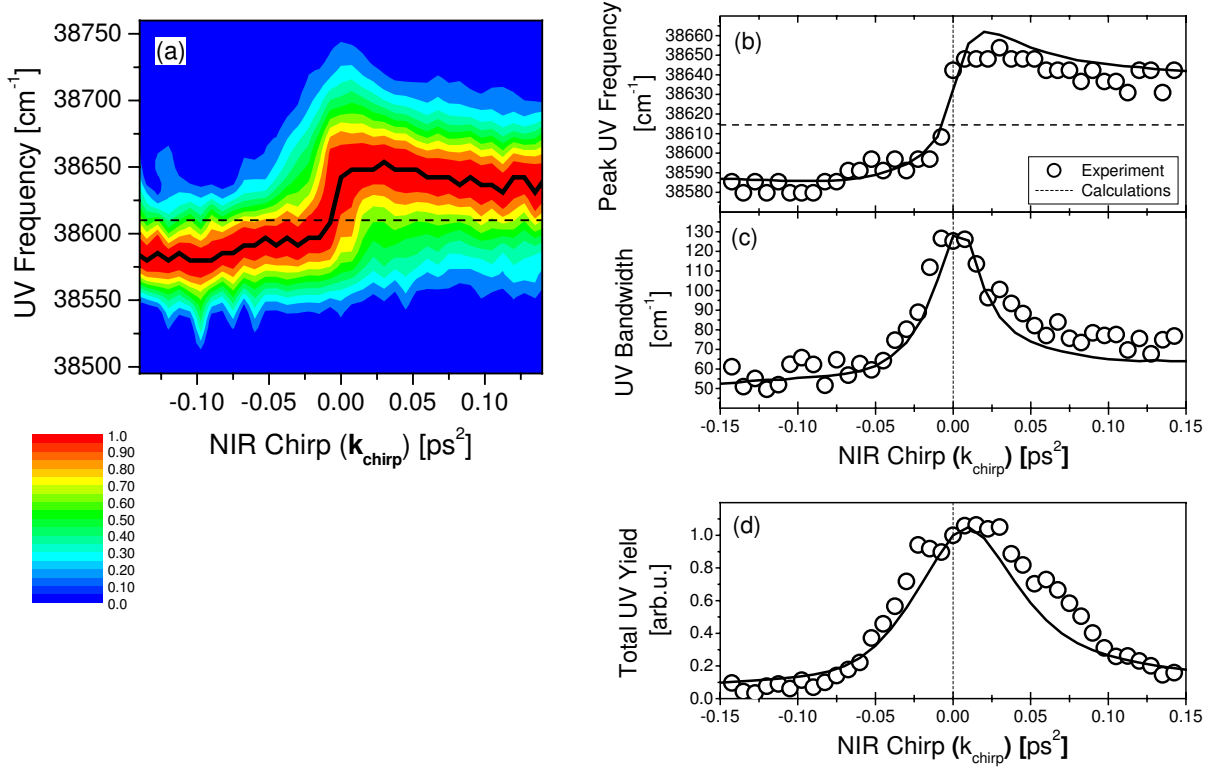}
\caption{(Color online) Results for controlling the multiphoton
resonance-mediated generation of coherent broadband UV radiation in
Na by linearly-chirped phase-shaped NIR femtosecond pulses. (a)
Experimental results for the spectral intensity (color-coded)
measured at each UV frequency as a function of the applied NIR
chirp. Each UV spectrum at a given chirp value is normalized
(independently) by its maximal intensity;
(b,c) Experimental results (circles)
for the NIR-chirp dependence of the measured peak UV frequency and UV bandwidth (FWHM),
as obtained from the data of (a).
Corresponding calculated results (lines) are also shown.
The dashed lines in (a) and (b) indicate the UV frequency of 3$\times$$\omega_{4s,3s}/2$ (see text);
(d) Experimental (circles) and calculated (lines) results for
the NIR-chirp dependence of the total UV yield.
They are normalized by the yield generated by the NIR TL pulse.} 
\label{fig_3}
\end{figure}

Figure~\ref{fig_3}(a) presents the complete set of 
the experimentally generated chirp-controlled UV spectra.
The results are presented as a color-coded map of the spectral intensity $I_{out}(\Omega; \kappa_{chirp})$
measured at each UV frequency $\Omega$
as a function of the $\kappa_{chirp}$ value applied to the generating linearly-chirped NIR pulse.
Each UV spectrum at a particular value of $\kappa_{chirp}$ is normalized (independently)
by its maximal intensity. 
The experimental chirp dependence 
of the peak frequency
(i.e., the frequency of maximal intensity) 
and bandwidth (FWHM) 
of the generated UV spectrum,
as obtained from the map data,
are shown in Figs.~\ref{fig_3}(b) and \ref{fig_3}(c) (circles). 
Also shown there are 
corresponding theoretical results (lines), which are
the outcome of numerical calculations based on Eqs.(\ref{eq1})-(\ref{eq4})
that have been followed by a convolution with the UV spectral measurement resolution.
As seen, the theoretical results describe very well the experimental ones. 

As $\kappa_{chirp}$ is experimentally tuned from negative to positive values,
the peak UV frequency $\Omega_{0}$ changes
from around 38580$-$38585~cm$^{-1}$ when $\kappa_{chirp}$$\le$$-$0.10~ps$^{2}$ 
to   around 38635$-$38640~cm$^{-1}$ when $\kappa_{chirp}$$\ge$0.10~ps$^{2}$, 
with a value of 38642~cm$^{-1}$ when $\kappa_{chirp}$=0. 
The change of $\Omega_{0}$ takes place mostly at the region of small chirp values around zero chirp.
A line of reference,
which we refer to further below,
is indicated in Figs.~\ref{fig_3}(a)-\ref{fig_3}(b) at
the UV frequency of 3$\times$$\omega_{4s,3s}/2$ (38610~cm$^{-1}$). 
The measured UV bandwidth
$\Delta\Omega$ amounts to about 125~cm$^{-1}$ around zero chirp 
and decreases continuously down-to 55$-$75~cm$^{-1}$ at the larger negative and positive chirps of $|\kappa_{chirp}|$$\ge$0.1~ps$^{2}$. 
After deconvoluting the experimental resolution,
the corresponding values are about 120~cm$^{-1}$ 
and 40$-$65~cm$^{-1}$, respectively. 
The chirp dependence of $\Delta\Omega$ 
is nearly symmetrical around $\kappa_{chirp}$=0. 
%
As explained above, all these specific chirp-dependent values of $\Omega_{0}$ and $\Delta\Omega$ and their tunability range
result from the specific NIR excitation spectrum used here. 
%

Highly-intuitive qualitative understanding of these chirp dependencies 
is actually obtained
by combining the frequency-domain picture of Eqs.~(\ref{eq1})-(\ref{eq4})
with the picture of the so-called
joint time-frequency domain.
Under this framework, 
zero chirp (TL pulse) corresponds to the simultaneous arrival of all the spectral pulse frequencies,
while positive and negative chirps correspond to the cases where lower frequencies arrive, respectively,
before or after higher frequencies.
The temporal separation between two different frequencies increases as the chirp magnitude increases.
With regard to 
the Na excitation,
the resonance-mediated (2+1) three-photon excitation to a given total energy of $\Omega$ 
is expected to be of large amplitude, leading to high UV intensity at $\Omega$,
once the third-photon absorption occurs 
shortly after 
the on- or near-resonant excitation of the intermediate
$4s$ state.
%
Based on Eq.~(\ref{eq4}) of $A^{(2)}$($\xi$$\approx$$\omega_{4s,3s}$),
such $4s$ excitation occurs by absorbing
all the NIR photon pairs having their frequency-sum close to $\omega_{4s,3s}$, 
i.e., their frequencies are symmetric around~$\sim$$\omega_{4s,3s}$/2, 
%
and thus can be 
associated with a narrow time window around the arrival time
of the pulse frequency $\omega_{4s,3s}$/2. 
Hence, overall,
efficient resonance-mediated excitation is expected
for the energies of $\Omega \approx \omega_{4s,3s}+\omega_{3}$
with $\omega_{3}$ 
either larger or smaller than $\omega_{4s,3s}/2$ according to whether the NIR chirp is positive or negative, respectively.
%
%
Consequently, the positive and negative NIR chirps are expected to lead
to efficient UV generation mostly at UV frequencies
that are either larger or smaller than 3$\times$$\omega_{4s,3s}/2$, respectively.
Also,
since larger chirp magnitude leads to larger temporal spread of the pulse frequencies
such that less frequencies arrive within the time window of efficient excitation, 
the generated UV bandwidth is expected
to be the largest for NIR chirps around zero and to decrease toward a constant level as the chirp magnitude increases.
Indeed, as seen in Figs.~\ref{fig_3}(a)-(c),
the observed chirp-dependent 
results 
follow this qualitatively-predicated behavior.

Last, for completing the picture, Fig.~\ref{fig_3}(d) presents
experimental (circles) and theoretical-numerical (lines)
results for the chirp dependence of the 
total generated UV yield $Y_{out}$.
The results are 
normalized by the yield 
generated by the NIR TL pulse. 
%
%
As $\kappa_{chirp}$ changes, the total UV yield 
experimentally changes continuously
from about 10\% to 105\% of the TL-generated UV yield.
%
From an applicative point of view, 
such a change in $Y_{out}$ is not optimal for using 
the generated shaped short-wavelength broadband radiation.
Nevertheless, having in hand a corresponding calibration curve, such as the one shown in Fig.~\ref{fig_3}(d),
removes this inconvenience.
Also here, similar to the above 
spectral UV results, 
the experimental results of $Y_{out}$ are described very well by the theoretical~ones.

In conclusion, we have introduced a new scheme for generating short-wavelength
coherent broadband radiation with well-controlled spectral characteristics.
It utilizes shaped long-wavelength femtosecond pulses to coherently control
atomic resonance-mediated multiphoton excitation to a broad far-from-resonance continuum.
As compared to a completely nonresonant atomic excitation,
the resonance-mediated nature of the excitation 
provides a higher degree of control over the properties of
the generated radiation 
together with a higher generation efficiency.
In the present work,  
by tuning the linear chirp applied to the exciting phase-shaped NIR femtosecond pulse
we have experimentally controlled the spectrum of the coherent broadband UV radiation
generated 
via resonance-mediated (2+1) three-photon excitation of Na.
The 
coherent UV bandwidth corresponds to a femtosecond UV pulse. 
This is a first step in implementing the full scheme for producing shaped femtosecond pulses
at wavelengths down-to the VUV range, which currently is inaccessible by 
the present 
pulse shaping techniques.
Inspired by the past demonstrations of narrowband coherent VUV generation (see above),
the extension to the VUV range will be achieved by a proper change of the 
excited atomic species and the driving pulse wavelength  
and/or by utilizing resonance-mediated (N+1) excitations of higher order (i.e., N$>$3).
%
%
Such shaped pulses will make many VUV-absorbing molecular species newly available for coherent control.
%
This is important 
for nonlinear spectroscopy and microscopy as well as
for chemical reaction photo-control. 



\newpage

\end{document}